\documentclass[a4paper,11pt]{article}
\usepackage{graphicx}
\usepackage{amsfonts}
\usepackage{amsmath}
\usepackage{hyperref}
\title{The (in)security of some recently proposed lightweight key distribution schemes}
\author{Chris J. Mitchell\\Information Security Group, Royal Holloway, University of London\\
\url{www.chrismitchell.net}}
\date{13th March 2021 (v3)}
\newcommand{\qed}{\nobreak \ifvmode \relax \else
      \ifdim\lastskip<1.5em \hskip-\lastskip
      \hskip1.5em plus0em minus0.5em \fi \nobreak
      \vrule height0.75em width0.5em depth0.25em\fi}
\begin{document}

\maketitle

\begin{abstract}
Two recently published papers propose some very simple key distribution schemes designed to enable
two or more parties to establish a shared secret key with the aid of a third party. Unfortunately,
as we show, most of the schemes are inherently insecure and all are incompletely specified ---
moreover, claims that the schemes are inherently lightweight are shown to be highly misleading.  We
also briefly critique a somewhat related very recent paper by the same authors that uses similar
techniques to achieve what are claimed to be secure multiparty computations.
\end{abstract}

\section{Introduction} \label{Intro}

Two papers recently published by Harn et al.\ \cite{Harn20a,Harn20} propose some very simple key
distribution schemes designed to enable two or more parties to establish a shared secret key with
the aid of a third party. Unfortunately, as we show, most of the schemes are inherently insecure
and all are incompletely specified.  Moreover, claims regarding the efficiency of the schemes,
namely that they are inherently lightweight, are shown to be highly misleading.  We also briefly
critically examine a related recent paper by the same authors \cite{Harn21}, which also has a
number of significant shortcomings.

The remainder of this paper is arranged as follows.  Section~\ref{sec:background} provides an
introduction to the protocols in question, and sets them in the context of the extensive prior art.
This is followed by analyses of the various proposed protocols: Sections~\ref{sec:A1},
\ref{sec:A2}, \ref{sec:A3} and \ref{sec:A4} provide analyses of the four protocols in the first
paper \cite{Harn20}, and Section~\ref{sec:B} briefly examines the almost identical protocols in the
second paper \cite{Harn20a}. Section~\ref{sec:SMPC} introduces and evaluates the very recently
published secure multiparty computations paper by the same set of authors. Finally,
Section~\ref{sec:Conclusions} concludes the paper.

\section{Background} \label{sec:background}

The two key distribution papers, whilst having very closely-related content and the same set of
authors, do not refer to each other, and it is hard to know which was submitted first. For
simplicity we refer throughout to \emph{Paper A} \cite{Harn20}, and \emph{Paper B} \cite{Harn20a}.
Since the two schemes in Paper B \cite{Harn20a} both appear to be slightly elaborated versions of
two of the schemes in Paper A \cite{Harn20}, we consider Paper A first.

\subsection{Paper A}

Paper A \cite{Harn20} presents a total of four closely related schemes, which we refer to below as
A1--A4. They all involve using a combination of a Key Derivation Function (KDF) (see, for example,
ISO/IEC 11770-6:2016 \cite{ISO11770-6:2016}) and pre-established shared secret keys to establish
new one-time session keys for pairs or groups of participants.

All the schemes work in essentially the same way.  Long-term secret keys are first used to create
one-time keys, using a KDF.  The newly generated one-time keys are then combined using bit-wise
exclusive-or to create the desired session keys.  We provide details of the individual schemes
below.

\subsection{Paper B}

Paper B \cite{Harn20a} presents two schemes which also reply on KDFs and pre-established shared
secrets --- we refer to these as B1 and B2. As we observe below, both B1 and B2 are essentially the
same as scheme A3.

\subsection{Threat models}  \label{subsec:threat_model}

Neither of the two papers explicitly give the threat model in the context of which the protocols
are designed to operate.  We therefore assume (as is common practice --- see, for example, Boyd et
al.\ \cite{Boyd20}) that messages are sent between participants via an unreliable channel, i.e.\
one where an adversary can eavesdrop on messages and also delete, modify or insert spurious
messages. Of course, any such channel can be upgraded to a secure channel using well-established
technology such as TLS \cite{RFC8446}, but in such a case the protocols described in the two papers
would not be needed --- keys could simply be transferred in cleartext.

\subsection{The prior art}

There is a large and very well-established literature on protocols designed to enable two parties
to establish a shared secret session key with the aid of a trusted third party with whom they both
share a long-term secret key. A classic example is that of Kerberos, whose design dates back to the
mid-1980s \cite{RFC1510,Kohl90,Steiner88}. An extensive discussion of such protocols can be found
in the landmark 1990s Handbook of Applied Cryptography \cite{Menezes97}, and the first edition of a
standard for secret-key-based key establishment protocols, ISO/IEC 11770-2, was published in 1996
\cite{ISO11770-2:1996}
--- the current version of the standard was published in 2018 \cite{ISO11770-2:2018}.  For an
up-to-date summary of the state of the art, and a review of the history of the subject area, the reader
is referred to Chapter 3 of the excellent Boyd et al.\ \cite{Boyd20}.

Unfortunately, much of this prior art is either not known to the authors or has been ignored.
Certainly, no attempt has been made in either of Papers A or B to provide a comparison of the
efficiency and/or the security properties of the proposed schemes with the prior art.

\section{Scheme A1} \label{sec:A1}

\subsection{Operation}

The first scheme in Paper A (\cite{Harn20}, \S 3.1) is designed to enable two parties ($A$ and $B$)
to establish a shared secret session key, with the aid of a trusted third party $C$.  $A$ and $B$
both share a long-term secret key with $C$ (which we label $L_{AC}$ and $L_{BC}$, respectively),
although $A$ and $B$ do not, a priori, share a key --- explaining the role of $C$.  Third party $C$
is trusted to the extent that it chooses the value of the secret session key to be shared by $A$
and $B$.

All parties are also assumed to have access to a KDF $f$, the nature of which is not precisely
specified.  For the purposes of this discussion, and in line with common practice, we suppose that
$f$ takes two inputs, namely (a) a secret key, and (b) some one-time data (which may be public),
and outputs a secret key (this is what is referred to in ISO/IEC 11770-6 as a \emph{key expansion
function}). The nature of this function is also not specified, but typically it is instantiated as
a CBC-MAC, i.e.\ a Message Authentication Code based on a block cipher in CBC mode (see, for
example, ISO/IEC 9797-1 \cite{ISO9797-1:2011}). It could also be instantiated using a cryptographic
hash function, although the computational complexity is likely to be very similar to use of a block
cipher.  Most importantly, regardless of how it is implemented, it must be computationally
infeasible to recover the input secret key even if a number of outputs are known. Note that all the
schemes in both papers use such a function $f$, and we implicitly assume throughout that this has
been agreed in advance.

The scheme has two phases.
\begin{enumerate}
\item In Phase 1, which is only briefly sketched in \S3.1.2 of Paper A:
\begin{itemize}
\item $A$, $B$ and $C$ (by some unspecified means) agree on a public nonce value $N$;
\item $A$ and $C$ use a public nonce $N$ and the KDF $f$ to compute a one-time shared
    secret key as $K_{AC}=f(L_{AC},N)$;
\item $B$ and $C$ make an analogous calculation to compute a one-time shared secret key as
    $K_{BC}=f(L_{BC},N)$.
\end{itemize}
It is implicit that $N$ should be generated in such a way that it is only ever used once.

\item In Phase 2, $C$ chooses a one-time session key $K_{AB}$ to be shared by $A$ and $B$,
    which must have the same bit-length as $K_{AC}$ and $K_{BC}$, and distributes it to them
    both in the following way:
    \begin{itemize}
    \item $C$ computes $C_A=K_{AB}\oplus K_{AC}$ (where here and throughout $\oplus$
        denotes bit-wise exclusive-or) and sends this value to $A$;
    \item $A$ computes $C_A\oplus K_{AC}=K_{AB}$ to recover the session key;
    \item in parallel, $C$ computes and sends $C_B=K_{AB}\oplus K_{BC}$ to $B$, who can
        compute $C_B\oplus K_{BC}=K_{AB}$.
    \end{itemize}
That is, at the end of Phase 2, $A$ and $B$ will share the session key $K_{AB}$.

\end{enumerate}

\subsection{Efficiency and comparisons}

The authors of Paper A make the following claims (see \cite{Harn20} \S3.1.2).
\begin{quote}
The operation of this proposed scheme is lightweight since it only needs to evaluate the positions
of matching bits between two keys which is equivalent to the computation of a logical exclusive or
(XOR) operation. In summary, the scheme is very efficient in computation since logic XOR is the
simplest operation and it is also efficient in communication since it is non-interactive.
\end{quote}
It is true that Phase 2 of the scheme does involve bit-wise exclusive-or operations.  However,
Phase 1 requires both $A$ and $B$ to compute the function $f$, i.e.\ to perform at least one block
cipher encryption or hash function computation, which is nowhere near so lightweight as an XOR ---
typically, a hash function computation is of the same order of complexity as a block cipher
encryption. The claim made in the paper is thus extremely misleading.

It is helpful to consider how the scheme compares with various well-established protocols of this
general type, i.e.\ in which a shared secret session key is established between two parties with
the help of a trusted third party, with whom both parties share a long-term secret key.  ISO/IEC
11770-2 \cite{ISO11770-2:2018} specifies four protocols of this type, namely Mechanisms 7--10,
which are referred to as \emph{Mechanisms using a Key Distribution Centre}. These protocols have
been extensively analysed over the 25 years since the standard first appeared, and (after some
minor modifications) there is now robust evidence that they meet the claimed security properties
(see, in particular, Cremers and Horvat \cite{Cremers14,Cremers16}).

The four protocols have slightly varying security properties, and two of them also include an
optional step designed to enable $A$ and or $B$ to confirm that their counterpart has successfully
received the shared key.  Since such a property is not provided by scheme A1, we ignore the
optional steps. Table B.1 in Annex B of ISO/IEC 11770-2 \cite{ISO11770-2:2018} helpfully provides a
detailed comparison of the protocols, stating that all four protocols require $A$ and $B$ to
perform one encryption operation, with the exception of Mechanism 10 which requires one of the two
parties to perform two such operations.  A superficial analysis therefore suggests that the
complexity of these existing standardised protocols is actually directly comparable with Scheme A1.

Finally note that, whilst I am not aware that the precise protocol has previously been proposed
(there are so many possible variants), something rather similar was proposed by Gong over 30 years
ago \cite{Gong89}. It is interesting to note that an issue was found with this latter scheme by
Boyd and Mathuria in 1997 \cite{Boyd97}.

\subsection{Security analysis}

Unfortunately it appears that Scheme A1 is subject to a simple attack.  We sketch one possible
attack scenario.  The attack hinges on the fact that there is no specified means for the parties to
agree on who is involved in the protocol at the beginning of Phase 1.  We assume that the protocol
set-up (including setting up the nonce) takes place via a public channel, with no protection for
the exchanged messages; of course, such an exchange could be made secure but this would require
further use of cryptography, making the protocol even less `lightweight'.

In line with the assumed threat model stated in Section~\ref{subsec:threat_model}, we suppose that
a malicious third party $E$ can control the communications channel with respect to a victim user
$A$. We suppose that $E$ convinces $A$ that a secret key is being established with party $B$ (who
is actually not involved).  The third party $C$ believes it is establishing a secret key between
$A$ and another (legitimate) party $D$.  All active parties (i.e.\ $A$, $C$ and $D$) agree on the
nonce $N$.

$C$ computes a one-time key for $A$ as $K_{AC}=f(L_{AC},N)$, chooses the session key $K_{AD}$ and
sends $K_{AC}\oplus K_{AD}$ to $A$ (the malicious party $E$ does not interfere with this message).
$A$ can now also compute $K_{AC}$ and uses it to recover $K_{AD}$, which $A$ believes is shared
with $B$.  Simultaneously $C$ computes a one-time key for $D$ as $K_{AD}=f(L_{AC},N)$, and computes
and sends $K_{DC}\oplus K_{AD}$ to $D$; $D$ now recovers $K_{AD}$, which $D$ (correctly) believes
is shared with $A$.

That is, at the end of the protocol $A$ has a session key which it believes is shared with $B$ but
is actually shared with $D$. This is clearly a breach of the protocol objectives, especially if we
observe that $D$ could be the same as the malicious entity $E$. Indeed, by repeating this deception
with party $B$, $E$ could end up with session keys shared with $A$ and $B$, which $A$ and $B$
believe are shared by each other.  This would enable $E$ to act as an eavesdropping
`man-in-the-middle', reading messages sent between $A$ and $B$ which $A$ and $B$ believe are
securely encrypted.

There are a range of possible `fixes' to the protocol. One possibility would be to include the
identifiers for $A$ and $B$ in the inputs to the KDF $f$, i.e.\ so that one-time keys are computed
as function of both a nonce and unique identifiers for the parties involved. However, as is
well-known, proposing ad hoc fixes to protocols without providing rigorous evidence of security is
a dangerous path, and so even a fixed-up version should only be considered for real-world use
subject to the recommendations below.

\subsection{Recommendations}

If such a protocol is to be seriously considered for use then the following points need to be
addressed.
\begin{itemize}
\item A modification to address possible attacks, such as that outlined above, needs to be
    fully specified.
\item A threat model needs to be carefully defined, and a formal security model for the
    properties of the protocol needs to be given that matches the threat model.
\item A proof of security in the chosen security model needs to be given.
\item A detailed performance and security comparison with existing protocols, such as those
    specified in ISO/IEC 11770-2 \cite{ISO11770-2:2018} needs to be given, so that any
    performance advantages over the prior art can be verified.
\end{itemize}

\section{Scheme A2} \label{sec:A2}

\subsection{Operation}

This scheme is described in \S 3.2 of Paper A \cite{Harn20}.  It is even simpler than the first
scheme.  It simply describes how an entity $A$ can be equipped by a trusted party $S$ with a new
secret session key, assuming $A$ and $S$ share a long-term secret key, $L_{AS}$ say.

There are four steps (although the first two are only briefly sketched in \S 3.2.2 of Paper A).
\begin{enumerate}
\item $A$ and $S$ (by some unspecified means) agree on a public nonce value $N$;
\item $A$ and $S$ use the public nonce $N$ and the KDF $f$ to compute a one-time shared secret
    key as $K_{AS}=f(L_{AS},N)$;
\item $S$ chooses a one-time session key $K$ to be shared by $A$ and $S$, which must have the
    same bit-length as $K_{AS}$, computes $C_S=K_{AS}\oplus K$, and sends this value to $A$;
\item $A$ recovers $K=C_S\oplus K_{AS}$.
\end{enumerate}

\subsection{Efficiency and comparisons}

Analogously to Scheme A1, the authors of Paper A make the following claims (\cite{Harn20},
\S3.2.2).
\begin{quote}
In summary, the scheme is very efficient in computation since it needs only one logical XOR
operation for both server and user and is efficient in communication since it is non-interactive.
\end{quote}
Just as for Scheme A1, the protocol requires $A$ and $S$ to compute the function $f$, i.e.\ at
least one block cipher encryption or hash function computation, which is nowhere near so
lightweight as a single bit-wise XOR. The efficiency claim is thus again extremely misleading.

Interestingly, steps 1 and 2 of the mechanism are essentially identical to the steps in Key
Establishment Mechanism 1 of ISO/IEC 11770-2 \cite{ISO11770-2:2018}.

\subsection{Security analysis and recommendations}

As the standard makes clear, ISO/IEC 11770-2 Key establishment mechanism 1 (which corresponds to
steps 1 and 2 of Scheme A2) does not provide authentication of the one-time key --- i.e.\ of the
key $K_{AS}$ using the above notation.  This is because there is no means provided for $A$ and $S$
to securely agree on the value of the nonce $N$ --- moreover, $C_S$ is also not authenticated. As a
result, it follows that the session key $K$ established by Scheme A2 is not authenticated, i.e.\ it
could even be the case that this key is known to a party other than $A$ or $S$.

Therefore, unless additional mechanisms are put in place to guarantee the timeliness and origin of
the nonce $C$ and the value $C_S$, the mechanism should not be used.  A set of six mechanisms of
this general type with well-understood security properties are given in Clause 6 of ISO/IEC 11770-2
\cite{ISO11770-2:2018}, and depending on the context and precise security requirements, one of
these should be used in preference to Scheme A2; that is, of course, unless the communications
channel is already made secure by other means.

\section{Scheme A3} \label{sec:A3}

This slightly more complex scheme can be found in \S 3.3 of Paper A.  It is designed to enable a
group of three or more users to establish a shared session key using pre-established long-term
shared secret keys.  Two versions are given: first a scheme designed specifically for three
parties, and subsequently a scheme for an arbitrary-sized group.  We consider them in turn.

\subsection{The three-party scheme}

The scheme is a simple derivative of Scheme A1.  The only difference is that pre-established
long-term secret keys are assumed to be shared by every pair of the three parties (labeled $A$, $B$
and $C$). These are then combined with an agreed nonce using a KDF to generate three one-time
shared secret keys $K_{AB}$, $K_{AC}$ and $K_{BC}$ (shared by $A$ and $B$, $A$ and $C$, and $B$ and
$C$, respectively).

Any one of the parties, which we assume to be $A$ (without loss of generality), computes
$C_A=K_{AB}\oplus K_{AC}$ and sends it to both $B$ and $C$.  Both $B$ and $C$ can now recover the
pair of keys ($K_{AB}$, $K_{AC}$) --- which are also known to $A$ --- and a combination of these
two keys, e.g.\ using a KDF, forms the group key.

Since it is essentially the same as Scheme A1 it suffers from the same serious security issues.
Moreover, while Paper A makes the usual claim about the protocol being lightweight, its (hidden)
dependence on use of the KDF for every instance of the protocol means that this claim is highly
misleading.

\subsection{The multi-party version}

This is an elaboration of the three-party version, which is only very briefly sketched in \S3.3.2
of Paper A\@. One of the parties is appointed as the initiator, and somehow all of the $n$ parties
agree on the identity of the initiator and a nonce.  The initiator is also assumed to pre-share a
long-term secret key with all the other $n-1$ parties.  To start the protocol, one-time keys are
generated as a function of the nonce and these long-term shared keys; as a result the initiator
will share a one-time secret key with every other party.

The $n$ parties are assumed to be arranged in a binary tree, rooted at the initiator (at the `top'
of the tree). The initiator then engages in a series of instances of the three-party protocol,
systematically working `down' the tree, at each stage establishing a group key shared by a larger
set of participants. Finally, when the `bottom' of the tree is reached, all participants share a
single group key.

Since the three-party protocol suffers from the same major security issues as Scheme A1, as
discussed above, similar issues arise with the $n$-party version of the scheme.  Yet again the
authors make the highly misleading claim that `the scheme is very efficient in computation since it
requires only one logical XOR operation for each user'.

\section{Scheme A4} \label{sec:A4}

\subsection{Goals of scheme}

The final scheme is described in \S 3.4 of Paper A \cite{Harn20}. The scheme has the same primary
objective as the three-party version of Scheme A3, namely to establish a shared group key amongst
three parties ($A$, $B$ and $C$) who have pre-established long-term secret shared keys. However,
one difference is that this scheme is claimed to provide an \emph{authenticated} group key,
although what this means, and the precise threat model, are left unspecified.  However, it is
stated that the scheme `should be able to prevent any outside attacker to impersonate to be any
legitimate user and discover the group key'.

This implicitly means that the other three schemes are \emph{not} authenticated --- indeed, this is
clear from the analyses of Schemes A1--A3 above.  It could therefore be argued that to claim the
previous schemes are insecure is unfair, as they were not designed to be secure against active
attackers. However, the following points need to be taken into account when evaluating such an
argument.
\begin{itemize}
\item At no point in Paper A (or Paper B) is there any effort to make clear that the schemes
    A1--A3 (and B1 and B2) are insecure and should only be used when the communications
    channels are secure, e.g.\ using an underlying protocol such as TLS\@.  Indeed, the
    requirement for such an underlying security protocol would completely invalidate any claims
    about lightweight properties.  Moreover, the authors state the following in \S 3.1 of Paper
    A (this is the part of the paper describing Scheme A1 --- the `basic scheme' in the
    language used below).
    \begin{quote}
This basic scheme can be applied to many practical network applications. For example, in a
wireless sensor network, each sensor has been pre-loaded with a subset of keys before deploying
sensors to a geographical area in a random key distribution solution. After deploying these
sensors, two neighboring sensors intend to negotiate a common key between them to establish a
secure communication. In order to increase the probability of sharing an overlapping key in two
different subsets of sensors, researchers have proposed various solutions. Our proposed key
distribution scheme provides an alternative solution for this application. In our scheme, two
sensors with no pre-shared key can establish a one-time common key through a third sensor in
which both sensors have pre-shared keys with the third sensor separately.
    \end{quote}
This text clearly recommends use Scheme A1 directly, i.e.\ without any additional security
measures, over a wireless sensor network, where such networks are likely to be deployed in an
environment where interception and manipulation of data transmissions will be feasible.
\item Possible attacks on Scheme A4 are described below, which enables the value of an agreed
    group key to be changed, i.e.\ so that one participant ends up with a key different to that
    of the two other participants. That is, despite being described as `authenticated', the
    scheme does not offer significantly more protection against malicious adversaries than any
    of the other schemes.
\item It is far from clear what the authors mean by `authenticated' key establishment; it would
    seem (see below) that the only additional security property for Scheme A4 is that it can
    detect if $A$ and $B$ send inconsistent values to $C$.
\end{itemize}

\subsection{Operation}

As for the three-party variant of Scheme A3, it is assumed that pre-established long-term secret
keys are shared by every pair of the three parties. These are then combined with an agreed nonce
using a KDF to generate three one-time shared secret keys $K_{AB}$, $K_{AC}$ and $K_{BC}$ (shared
by $A$ and $B$, $A$ and $C$, and $B$ and $C$, respectively).  We suppose that the versions of the
keys held by $A$ are $K^1_{AB}$, $K^1_{AC}$, the versions of the keys held by $B$ are $K^2_{AB}$,
$K^2_{BC}$, and the version of the keys held by $C$ are $K^3_{AC}$, $K^3_{BC}$.  Of course if
everything is working correctly then $K^1_{AB}=K^2_{AB}$, $K^1_{AC}=K^3_{AC}$, $K^2_{BC}=K^3_{BC}$,
but the protocol is designed to detect such inconsistencies.

Two of the three parties now jointly act as protocol initiators (which, without loss of generality,
we assume are $A$ and $B$).  Both $A$ and $B$ now conduct the three-party version of Scheme A3,
which enables all parties to agree on the group key $K=K^1_{AB}=K^2_{AB}$.
\begin{itemize}
\item $A$ computes $C_A=K^1_{AB}\oplus K^1_{AC}$ (using its copies of these two keys) and sends
    it to $C$;
\item $B$ computes $C_B=K^2_{AB}\oplus K^2_{AC}$ (again using its copies of these two keys) and
    sends it to $C$;
\item $C$ computes $K^1=C_A\oplus K^3_{AC}$ and $K^2=C_B\oplus K^3_{BC}$.  If $K^1=K^2$, which
    should be true if all parties hold the same pre-shared keying material, the same nonce, and
    all messages are exchanged correctly, then $C$ accepts $K^1=K^2$ as the group key (as do
    $A$ and $B$).
\end{itemize}

\subsection{Analysis}

As for the three previous schemes, \S 3.4.2 of Paper A makes a claim about the efficiency of the
scheme.
\begin{quote}
Each initiator needs to execute one logic XOR operation to compute the positions of matching bits
between two pairwise shared keys and the third user needs to execute two XOR operations to extract
and compare a pairwise shared key which the user has no prior knowledge of it. The overall
performance of this scheme is lightweight.
\end{quote}
As in the previous cases, there is no mention of the cost of applying a KDF, which means that the
efficiency claims are highly misleading.

Even more seriously, the scheme is not significantly more secure than Schemes A1 and A3, as
somewhat analogous attacks apply.  We sketch two possible attack scenarios.  In both cases, in line
with the assumed threat model stated in Section~\ref{subsec:threat_model}, we suppose that a
malicious third party $E$ can control the communications channel with respect to a victim user $C$.

In the first attack, $E$ first chooses a block of bits $M$ of the same length as the key $K_{AB}$.
$E$ now modifies the messages sent from $A$ to $C$ and $B$ to $C$ as follows:
\begin{itemize}
\item $C_A$ is replaced with $C_A\oplus M$; and
\item $C_B$ is replaced with $C_B\oplus M$.
\end{itemize}
It is not hard to see that C will compute $K^1=K^2=K_{AB}\oplus M$.  That is, the value accepted by
$C$ will be different to the value held by $A$ and $B$, i.e.\ the method does not provide key
authentication as claimed.  Of course, this attack could probably be `fixed' by providing explicit
authentication for the messages containing $C_A$ and $C_B$, e.g.\ by adding a MAC to the two
messages. However, this would make the protocols much less efficient, voiding any justification for
departing from the well-established state of the art.

The second attack requires $E$ to be able to manipulate the nonce value agreed by the participants.
This is a plausible assumption --- there is certainly no discussion in either of the two papers on
a secure method for agreeing this nonce, and it is therefore reasonable to assume it is agreed by
insecure messaging (adding an exchange to agree the nonce in a secure way would essentially require
a security protocol to be run in advance of the one proposed).  By comparison, this issue \emph{is}
addressed in previously proposed nonce-based protocols, by making the nonce exchange an explicit
part of the protocol.  The attack proceeds in three stages.
\begin{enumerate}
\item Suppose $E$ convinces $A$ to run the protocol with itself and party $C$, where $A$ and
    $E$ are the initiators and $N$ is the nonce. Note that $C$ is not actually involved --- $E$
    can send to $A$ messages that are apparently from $C$ agreeing to use of the protocol, as
    required. Suppose also that $E$ intercepts the message $C_A=K_{AE}\oplus K_{AC}$ sent from
    $A$ to $C$ (and prevents it from reaching $C$)\@. Since $E$ knows $K_{AE}$, $E$ can now
    recover the value of $K_{AC}$ for this value of the nonce $N$.
\item In a very similar way, $E$ convinces $B$ to run the protocol with itself and party $C$,
    where $B$ and $E$ are the initiators and the value of $N$ is the nonce. Again $C$ is not
    actually involved, and as previously we suppose that $E$ intercepts the message
    $C_B=K_{BE}\oplus K_{BC}$ sent from $B$ to $C$ (and also as in the previous step prevents
    it from reaching $C$). Since $E$ knows $K_{BE}$, $E$ can recover the value of $K_{BC}$ for
    this value of the nonce $N$.
\item Finally, $E$ persuades $C$ to run the protocol with $A$ and $B$, where $A$ and $B$ are
    the initiators and the same value $N$ is the nonce. The two parties $A$ and $B$ are not
    actually involved, and $E$ send messages to $C$ that are apparently from $A$ and $B$, as
    necessary. $E$ now chooses an arbitrary key $K^*$ and sends both $C^*_A=K_{AC}\oplus K^*$
    to $C$ (impersonating $A$), and $C^*_B=K_{BC}\oplus K^*$ to $C$ (impersonating $B$), where
    $K_{AC}$ and $K_{BC}$ are the values learnt from the first two steps of the attack.  It is
    straightforward to see that $C$ will accept $K^*$ as a valid group key shared with $A$ and
    $B$, whereas in fact it is known to $E$ (and not to $A$ and $B$).
\end{enumerate}

It is important to note that no party is asked to use the same nonce value twice, so even
if all parties keep a record of all the nonce values they have used (and refuse to reuse a
value) the attack can still operate.

\section{Schemes B1 and B2} \label{sec:B}

The schemes are essentially the same as Scheme A3 in Paper A, with some slight variations in the
wording.  More precisely the schemes in Paper B have the following relationship to the schemes in
Paper A.
\begin{itemize}
\item Scheme B1 (given in \S 3 of Paper B) is precisely the same as the three-party version of
    Scheme A1.
\item Scheme B2 (described in \S 4 of Paper B) is a group key distribution scheme and comes in
    two variants.  The first variant (see \S 4.1 of Paper B) involves multiple uses of Scheme
    A1, in a somewhat inefficient way.  The second variant (\S 4.2) is the same as the
    $n$-party variant of Scheme A2, i.e.\ it involves use of a binary tree.
\end{itemize}

Just as in Paper A, claims are made about the efficiency of the schemes, completely ignoring the
fact that use of the scheme involves computing a KDF.

\section{A related paper}  \label{sec:SMPC}

\subsection{The schemes}

In this last main part of the paper we consider the multiparty computation schemes described by
Harn, Xia and Hsu \cite{Harn21}.  The paper describes three schemes, all of which will work for an
arbitrary number of parties but which are only explicitly given for three (or, in the third case,
four) parties.  We follow the same approach here, observing that the generalisation to the
multiparty case is clear.

All schemes have the same goal, namely to enable the evaluation of $d_A+d_B+d_C$ without parties
$A$, $B$ and $C$ divulging the values of $d_A$, $d_B$ and $d_C$, respectively, to any other party.
In all three schemes, the involved parties $A$, $B$ and $C$ (and $D$ in the third scheme) have
pre-established shared secret keys ($K_{XY}$ shared by parties $X$ and $Y$) which are only ever
used once, i.e.\ new shared keys must be established prior to every use of the scheme.

\begin{enumerate}
\item In the first scheme $A$ (who knows $d_A$) computes and broadcasts
    $\Delta_A=d_A+K_{AB}+K_{AC}$. Analogously, $B$ and $C$ compute and broadcast
    $\Delta_B=d_B-K_{AB}+K_{BC}$ and $\Delta_C=d_C-K_{AC}-K_{BC}$. As a result all three
    parties can compute $\Delta_A+\Delta_B+\Delta_C=d_A+d_B+d_C$; however, no information is
    divulged regarding $d_A$, $d_B$ or $d_C$. As the authors observe, the same calculation can
    be performed by any party who intercepts the broadcasts, and hence the scheme does not
    preserve confidentiality.
\item The second scheme is the same as the first except that $C$ does not broadcast $\Delta_C$.
    As a result, $C$ is the only party able to compute $d_A+d_B+d_C$, i.e.\ it provides
    confidentiality of the computed value.
\item The third scheme is somewhat similar to the second scheme and involves a fourth party
    $D$, who is the only party able to compute $d_A+d_B+d_C$. The computed and broadcast values
    in this case are $\Delta^*_A=\Delta_A+K_{AD}$, $\Delta^*_B=\Delta_B+K_{BD}$ and
    $\Delta^*_C=\Delta_C+K_{CD}$, and $D$ recovers $d_A+d_B+d_C$ as
    $\Delta^*_A+\Delta^*_B+\Delta^*_C-K_{AD}-K_{BD}-K_{CD}$.
\end{enumerate}

\subsection{Specification issues}

The use of addition as the operator in the above descriptions suggests that all values are to be
treated as integers, i.e.\ elements of $\mathbb{Z}$.  However, whilst the scheme would work after a
fashion in this scenario, it would also leak information about the values $d_A$, $d_B$ and $d_C$.
To see this, suppose that in the case of the first scheme all the keys $K_{XY}$ are chosen to lie
in the range $[0,N-1]$ for some large $N$. Then if it happens that $\Delta_A=d_A+K_{AB}+K_{AC}$
equals $2N+\epsilon$ for some positive $\epsilon$, then it immediately follows that
$d_A\geq\epsilon+2$, i.e.\ anyone listening to the broadcast can potentially learn information
about $d_A$.

It would seem likely that all the values ($d_A$, $d_B$ and $d_C$ and the keys $K_{XY}$) are
intended to be chosen from the range $[0,N-1]$ for some large $N$, and that all the additions and
subtractions are computed modulo $N$.  If the keys are chosen uniformly at random from $[0,N-1]$,
then the protocol should not leak information about the values $d_A$, $d_B$ and $d_C$; however,
again, no advice is given on the choice of the keys.

The above interpretation seems reasonable in the light of the fact that an alternative version of
the first scheme is described in which (a) addition is replaced by multiplication, and (b)
subtraction is replaced by multiplication by the multiplicative inverse.  This version would make
no sense if calculations were in $\mathbb{Z}$, since there are no multiplicative inverses in
$\mathbb{Z}$! Indeed, the most likely interpretation of the multiplicative version of the scheme is
for $N$ to be chosen to be a prime, so that every value has a multiplicative inverse.  More
generally, it would make far more sense for the scheme to be described in the context of an
arbitrary group (of suitably large size).

\subsection{Other shortcomings}

If we suppose that the schemes are specified in the context of an arbitrary group, then, although
the system now works, problems still remain.  We focus on two such issues.

\begin{itemize}
\item \emph{No authentication}. None of the schemes enable the participants to verify the
    origin or integrity of the broadcast messages. This issue is not addressed by the authors.
    Without origin/integrity protection, the scheme would appear to be unusable.  Of course,
    such properties could be provided by layering the protocol on top of another protocol that
    provides integrity and authenticity protection for transferred messages. However, this
    would significantly increase the complexity of computations, communications and key
    management, very probably rendering any claimed advantages over rival schemes invalid.
\item \emph{Output value control}. Whilst this may not be an issue in all applications, the way
    in which the first scheme is designed enables one party to control the final computed
    value, i.e.\ $d_A+d_B+d_C$.  The party wishing to do this simply waits for the other
    parties to broadcast their messages, and chooses their $d_X$ value to give the desired
    outcome. Such an attack will work regardless of the number of participants.
\end{itemize}

\section{Concluding remarks}  \label{sec:Conclusions}

We have analysed a range of key establishment schemes proposed in two recent papers. There are a
number of major issues with these schemes, which we now summarise.
\begin{itemize}
\item The schemes have a range of serious security weaknesses, which potentially allow active
    opponents of the protocols to learn keys that they should not --- this arises partly
    because at no point are the security assumptions on which the protocols rest made clear.
\item All the schemes are claimed to be highly efficient (`lightweight') since they rely on
    computation of bit-wise exclusive-or operations to obtain the session key. However, in no
    case is a reference made to the fact that all the schemes require all parties to perform at
    least one key derivation computation, the cost of which makes the schemes no more
    lightweight than many rival schemes based on symmetric cryptography.
\item Paper B is essentially a subset of paper A.  Paper A does not refer to Paper B and
    neither does Paper B refer to Paper A\@. It appears they were submitted in parallel,
    although it is difficult to be sure since the published version of Paper A does not
    disclose when the first version was submitted.
\end{itemize}

In conclusion, neither of the key distribution papers add anything significant to the literature,
especially as no attempt is made to compare the schemes to the prior art, provide a detailed
description of (let alone prove) their security properties, or give a fair description of their
computational complexity.  The issues raised by publishing the same material twice also merit
careful consideration.

Finally, the proposed use of a similar approach to achieve multiparty computation is also shown to
have significant shortcomings, not least that the specification is very incomplete and that the
results of performing the protocols are completely unauthenticated.


\begin{thebibliography}{10}

\bibitem{Boyd97} C.~Boyd and A.~Mathuria.
\newblock Systematic design of key establishment protocols based on one-way
  functions.
\newblock {\em IEE Proceedings --- Computers and Digital Techniques},
  144:93--99, March 1997.

\bibitem{Boyd20} C.~Boyd, A.~Mathuria, and D.~Stebila.
\newblock {\em Protocols for Authentication and Key Establishment}.
\newblock Information Security and Cryptography. Springer, 2nd edition, 2020.

\bibitem{Cremers14} C.~Cremers and M.~Horvat.
\newblock Improving the {ISO/IEC} 11770 standard for key management techniques.
\newblock In L.~Chen and C.~J. Mitchell, editors, {\em Security Standardisation
  Research --- First International Conference, {SSR} 2014, London, UK, December
  16-17, 2014. Proceedings}, volume 8893 of {\em Lecture Notes in Computer
  Science}, pages 215--235. Springer, 2014.

\bibitem{Cremers16} C.~Cremers and M.~Horvat.
\newblock Improving the {ISO/IEC} 11770 standard for key management techniques.
\newblock {\em Int. J. Inf. Sec.}, 15(6):659--673, 2016.

\bibitem{Gong89} L.~Gong.
\newblock Using one-way functions for authentication.
\newblock {\em ACM Computer Communication Review}, 19(5):8--11, October 1989.

\bibitem{Harn20a} L.~Harn, C.~Hsu, and Z.~Xia.
\newblock Lightweight group key distribution schemes based on pre-shared
  pairwise keys.
\newblock {\em {IET} Commun.}, 14(13):2162--2165, 2020.

\bibitem{Harn20} L.~Harn, C.~Hsu, and Z.~Xia.
\newblock Lightweight and flexible key distribution schemes for secure group
  communications.
\newblock {\em Wireless Networks}, 27:129--136, 2021.

\bibitem{Harn21} L.~Harn, Z.~Xia, and C.~Hsu.
\newblock Non-interactive secure multi-party arithmetic computations with
  confidentiality for {P2P} networks.
\newblock {\em Peer-to-Peer Netw. Appl.}, 14(2):722--728, 2021.

\bibitem{ISO11770-2:1996} International Organization for Standardization, Gen\`{e}ve, Switzerland.
\newblock {\em {ISO/IEC 11770-2:1996, Information technology --- Security
  techniques --- Key management --- Part 2: Mechanisms} using symmetric
  techniques}, 1996.

\bibitem{ISO9797-1:2011} International Organization for Standardization, Gen\`{e}ve, Switzerland.
\newblock {\em {ISO}/{IEC} 9797--1, {Information technology} --- {Security
  techniques} --- {Message Authentication Codes (MACs)} --- {Part 1: Mechanisms
  using a block cipher}}, 2nd edition, 2011.

\bibitem{ISO11770-6:2016} International Organization for Standardization, Gen\`{e}ve, Switzerland.
\newblock {\em {ISO/IEC 11770-6:2016}, {Information technology} --- {Security
  techniques} --- {Key management} --- {Part 6: Key derivation}}, 1st edition,
  2016.

\bibitem{ISO11770-2:2018} International Organization for Standardization, Gen\`{e}ve, Switzerland.
\newblock {\em {ISO/IEC 11770-2:2018, IT Security techniques --- Key management
  --- Part 2: Mechanisms} using symmetric techniques}, 3rd edition, 2018.

\bibitem{RFC1510} J.~Kohl and C.~Neuman.
\newblock {\em {RFC 1510, The Kerberos Network Authentication Service (V5)}}.
\newblock Internet Engineering Task Force, September 1993.
\newblock \url{https://tools.ietf.org/html/rfc1510}.

\bibitem{Kohl90} J.~T. Kohl.
\newblock The use of encryption in {K}erberos for network authentication.
\newblock In G.\ Brassard, editor, {\em Advances in Cryptology --- {CRYPTO}
  '89, 9th Annual International Cryptology Conference, Santa Barbara,
  California, USA, August 20--24, 1989, Proceedings}, volume 435 of {\em
  Lecture Notes in Computer Science}, pages 35--43. Springer-Verlag, Berlin,
  1990.

\bibitem{Menezes97} A.~J. Menezes, P.~C. van Oorschot, and S.~A. Vanstone.
\newblock {\em Handbook of Applied Cryptography}.
\newblock CRC Press, Boca Raton, 1997.

\bibitem{RFC8446} E.~Rescorla.
\newblock {\em {RFC} 8446, {The Transport Layer Security (TLS) Protocol
  Version} 1.3}.
\newblock Internet Engineering Task Force, August 2018.
\newblock \url{https://tools.ietf.org/html/rfc8446}.

\bibitem{Steiner88} J.G.\ Steiner, C.\ Neuman, and J.I.\ Schiller.
\newblock Kerberos: an authentication service for open network systems.
\newblock In {\em Proceedings: Usenix Association, Winter Conference, Dallas
  1988}, pages 191--202. USENIX Association, Berkeley, California, February
  1988.

\end{thebibliography}

\end{document}